\documentclass{article}
\usepackage{ismir,amsmath,cite,url}
\usepackage{graphicx}
\usepackage{color}

\title{Large-Scale Cover Song Detection in Digital Music Libraries Using Metadata, Lyrics and Audio Features}

\oneauthor{Albin Correya, Romain Hennequin, Micka\"el Arcos} {Deezer R\&D, Paris\\research@deezer.com}
  
\begin{document}

\maketitle
\begin{abstract}
Cover song detection is a very relevant task in Music Information Retrieval (MIR) studies and has been mainly addressed using audio-based systems. Despite its potential impact in industrial contexts, low performances and lack of scalability have prevented such systems from being adopted in practice for large applications. In this work, we investigate whether textual music information (such as metadata and lyrics) can be used along with audio for large-scale cover identification problem in a wide digital music library. We benchmark this problem using standard text and state of the art audio similarity measures. Our studies shows that these methods can significantly increase the accuracy and scalability of cover detection systems on Million Song Dataset (MSD) and Second Hand Song (SHS) datasets. By only leveraging standard \textit{tf-idf} based text similarity measures on song titles and lyrics, we achieved 35.5{\%} of absolute increase in mean average precision compared to the current scalable audio content-based state of the art methods on MSD. These experimental results suggests that new methodologies can be encouraged among researchers to leverage and identify more sophisticated NLP-based techniques to improve current cover song identification systems in digital music libraries with metadata.
\end{abstract}
\section{Introduction}\label{sec:introduction}

From the advent of digital music platforms, music listening has been shifting from physical storage devices to digital music streaming platforms. Thousands of new songs are ingested daily to these platforms. Some music streaming services claims to have more than 40 million of songs in their catalogs. Hence, structuring and organizing songs in such large databases is of great interest since it can help to improve music recommendation and user-experience (to name a few). Beside that, understanding the underlying relations of a song and its cover versions are also beneficial for musicological studies. Cover song detection is a well identified task in MIR studies. It is typically set as a \textit{query-retrieval} task where the system retrieves  a ranked list of possible covers of a given query song from a reference database. Cover songs may have variations in key, tempo, structure, melody, harmony, timbre, language and lyrics compared to its original version, which makes it a challenging task. Over the past decade, numerous studies have been done on this problem using audio-based methods. An extensive review of traditional cover song detection system can be found in \cite{SerraReview2010}. Most of the best performing cover detection systems according to MIREX\footnote{http://www.music-ir.org/mirex/wiki/} framework rely on exposing the underlying tonal content of cover/non-cover pairs to compute the music similarities within them since most of the cover songs usually share considerable tonal similarity between each other.

Tonal features are mid-level audio features that correspond to the musical tonality of a sound such as key, melody, harmony or chords. Among them, chroma audio features have been widely used for the task of cover identification since it is a timbre invariant and octave-independent representation of the note-based tonal progression of a song over time. Variants of chroma features such as Harmonic Pitch Class Profile (HPCP) \cite{Serra2008:06,Serra2009:05} and Chroma Energy Normalized Statistics (CENS) \cite{MuellerChromaToolbox_ISMIR} were commonly used by existing systems since it gives an added robustness to tuning deviation, noise and local tempo variations compared to a normal chroma feature. Beside chroma features, melody (pitch salience) \cite{salamon2013tonal}, chord profiles \cite{KhadkevichChord2013:07}, self-similarity MFCC matrix \cite{tralie2017mfcc}, cognition-inspired descriptors \cite{van2014cognition} were also used in the literature. A cover song similarity distance is then computed using various sequence alignment techniques such as dynamic time warping \cite{Serra2008:06}, smith-waterman algorithm \cite{tralie2017mfcc}, recurrence quantification analysis \cite{Serra2009:05}, cross-correlation plots \cite{Zellis2006beatsync}, SiMPle \cite{silva2016simple} and  information-theoretic measures \cite{foster2015identifying}. A preprocessing step is usually done to these tonal features before computing this distance in order to make them invariant to the key  or the tempo of the song. Key invariance can be obtained by Optimal Transposition Index (OTI) \cite{OTISerra} or by computing 2D-Fourier transform magnitude coefficients of tonal feature vectors by disregarding the phase coefficients \cite{BertinMahieux2012:01}. Tempo invariance can be achieved by using beat-synchronous chroma features\cite{Zellis2006beatsync, tralie2017mfcc}.

On the other hand, machine learning techniques have been less explored in solving this task. In 2009, Ravuri et. al \cite{Ravuri2009} used a SVM classifier fed with different distance measures computed from chroma features at various tempos to classify cover/non-cover pairs (best second performing system in MIREX). This idea of combining similarity measures from different features was also further explored in the recent works of \cite{Osmalskyj2015:08,tralie2017mfcc,XXchen2018fusing}. In \cite{HeoMetricLearning17}, the authors explored the use of metric learning by projecting audio features in a high dimensional space where simple distance measures capture cover song similarity. Recently in \cite{CNNCover2017audio}, the authors trained a CNN with cross similarity matrices obtained from CENS features of cover/non-cover pairs to predict the probability of whether a reference song is a cover or non-cover of the query song. The idea of efficient database pruning and thus achieving multi-stage architecture for cover identification was explored in the works of \cite{EfficientpruningOsmalski} and \cite{Cai2017TwolayerCover}. Results reported in MIREX evaluations show little to no improvements for cover detection since 2009. The system proposed in \cite{Serra2009:05, Serra2008:06} still remains as the best performing algorithms in MIREX with a mean average precision (MAP) of $0.75$ on the MIREX mixed music collections. However MIREX cover detection evaluation is done on a dataset of 1000 songs with pairwise comparisons systems that are not scalable. To address this issue, several researchers studied and formulated the problem of large-scale cover identification. But the accuracy of these systems were not satisfactory (the best MAP reported is $0.28$ on the Second Hand Song dataset (SHS) train set alone \cite{Humphrey2013:02}). A more detail review of scalable cover detection systems can be found in Section \ref{sec:large-scale}. 
Recently there has been a growing interest in applying domain-specific knowledge such as from NLP on solving traditional MIR tasks using multi-modal approaches such as in genre classification \cite{OramasMulti-Label2017} and music recommendation \cite{OramasRecSys2017}. Even though NLP techniques were previously used in MIR studies for different tasks such as lyrics-alignment \cite{knees2005Lyrics}, mood, tag, genre classification  \cite{SchedlReview2016music}, particularly for the cover detection problem, the reliability and the accuracy of text-based approaches have not been studied or benchmarked beside in the recent work of \cite{SmithYoutube2017:00}. In this work, our main purpose is to benchmark the problem of large-scale cover identification using multi-modal data such as song title, lyrics and audio in a general framework considering all these different modalities of data are available to us. We believe that the results presented in this work could be a new direction for future industrial-scale cover identification systems dealing with commercial digital music libraries where textual, content and semantic informations of music are mostly available. The remainder of this paper is organized as follows: we describe the problem of large-scale cover identification, the related methodologies, datasets and evaluation frameworks in Section \ref{sec:large-scale}. In Section \ref{sec:Experiments}, we explain the setup of our experimental system. In Section \ref{sec:final-eval}, we discuss  and evaluate the proposed system in comparison to the state-of-the-art. Finally, we draw some conclusions in Section \ref{sec:conclusion}.

\section{Large-scale cover identification}\label{sec:large-scale}

\subsection{Related works}

Acknowledging the limits in terms of scalability and tractability of traditional cover detection systems, Bertin-Mahieux et. al \cite{BertinMahieux2011:03} used audio fingerprinting-inspired hashing to compute jump-codes from beat-synchronous chroma \cite{Zellis2006beatsync} to detect covers. Audio fingerprinting \cite{canofingerprintreview} methods were successfully used in the industry to find an almost identical copies of a given piece of music with high robustness to pitch, noise, distortions and other transformations. Later, audio fingerprinting strategies were explored in the works of \cite{RafiLiveFingerprint, KnownArtistLive} for detecting live version of cover songs. Authors of \cite{BertinMahieux2012:01} and  \cite{Humphrey2013:02} tried to achieve the factor of scalability and invariance by efficiently capturing some of this invariance in a low-dimensional feature space using sparse representations where simple distance metrics give a notion of cover song similarity. \cite{Humphrey2013:02} reported highest MAP score on the large-scale cover identification problem. Note that this is different to the non-scalable MIREX based evaluations we discussed earlier. Another way to achieve scalability in cover detection systems is by doing efficient database-pruning using various features in a preliminary step. This enabled us to use audio algorithms with more specificity to capture more understandable cover similarity in the later layers. This idea of multi-stage architecture were explored in the works of \cite{SmithYoutube2017:00, Cai2017TwolayerCover, Osmalskyj2015:08, EfficientpruningOsmalski} using audio based methods.

\subsection{Datasets}

As we talk about the large-scale cover identification problem, most of the public datasets like covers80\footnote{https://labrosa.ee.columbia.edu/projects/coversongs/covers80/} and covers1000\footnote{http://www.covers1000.net/} are inadequate since it has only a few hundred or thousands of songs. Identifying this problem, authors of \cite{BertinMahieux2012:01} also proposed the SHS dataset, a subset of MSD \cite{MSD2011} which has almost eighteen thousand songs clustered as cover cliques. Since it is a subset of MSD, the entire MSD can be used to evaluate cover identification systems and thus make it comparable to commercial scale applications. So far from our knowledge, SHS and MSD are the only publicly available dataset for large-scale cover identification tasks. The Musixmatch (MXM)\footnote{https://labrosa.ee.columbia.edu/millionsong/musixmatch} dataset provides \textit{bag-of-words} (BOW) lyrics for 23.76{\%}  songs in the MSD. Table \ref{tab:datasets} shows the summary of datasets used throughout our experiments. Among them, the mapping of MSD and Deezer audio database (MSD-DZR) were internally available from Deezer\footnote{http://www.deezer.com} which enabled us to run the audio-based experiments. We further explain various existing and observed problems in these datasets.

\begin{table}
\begin{center}
 \begin{tabular}{|p{1.6cm}|p{5.4cm}|}
  \hline
  Dataset & Description \\
  \hline
  MSD &  Metadata and precomputed audio features for 1 million songs\\
  \hline
  SHS & 18,196 MSD songs organized as cover cliques (available as train {\&} test splits)\\
  \hline
  Musixmatch (MXM)  &  Mapping of \textit{bag-of-words} lyrics provided by Musixmatch for 237,662 songs of MSD\\
  \hline
  MSD-DZR & Mapping of 82.45{\%} MSD songs with the Deezer audio database \\
  \hline
  SHS-DZR & Subset of MSD-DZR with corresponding mapping to SHS (with train and test splits) \\
  \hline
 \end{tabular}
 \end{center}
 \caption{Summary of datasets used in this study.}
 \label{tab:datasets}
\end{table}

\subsubsection{Similarities within cover set}\label{sec:cover-set-similarity}

\begin{figure}
 \centerline{
 {
 \includegraphics[width=8.4cm,height=5.4cm]{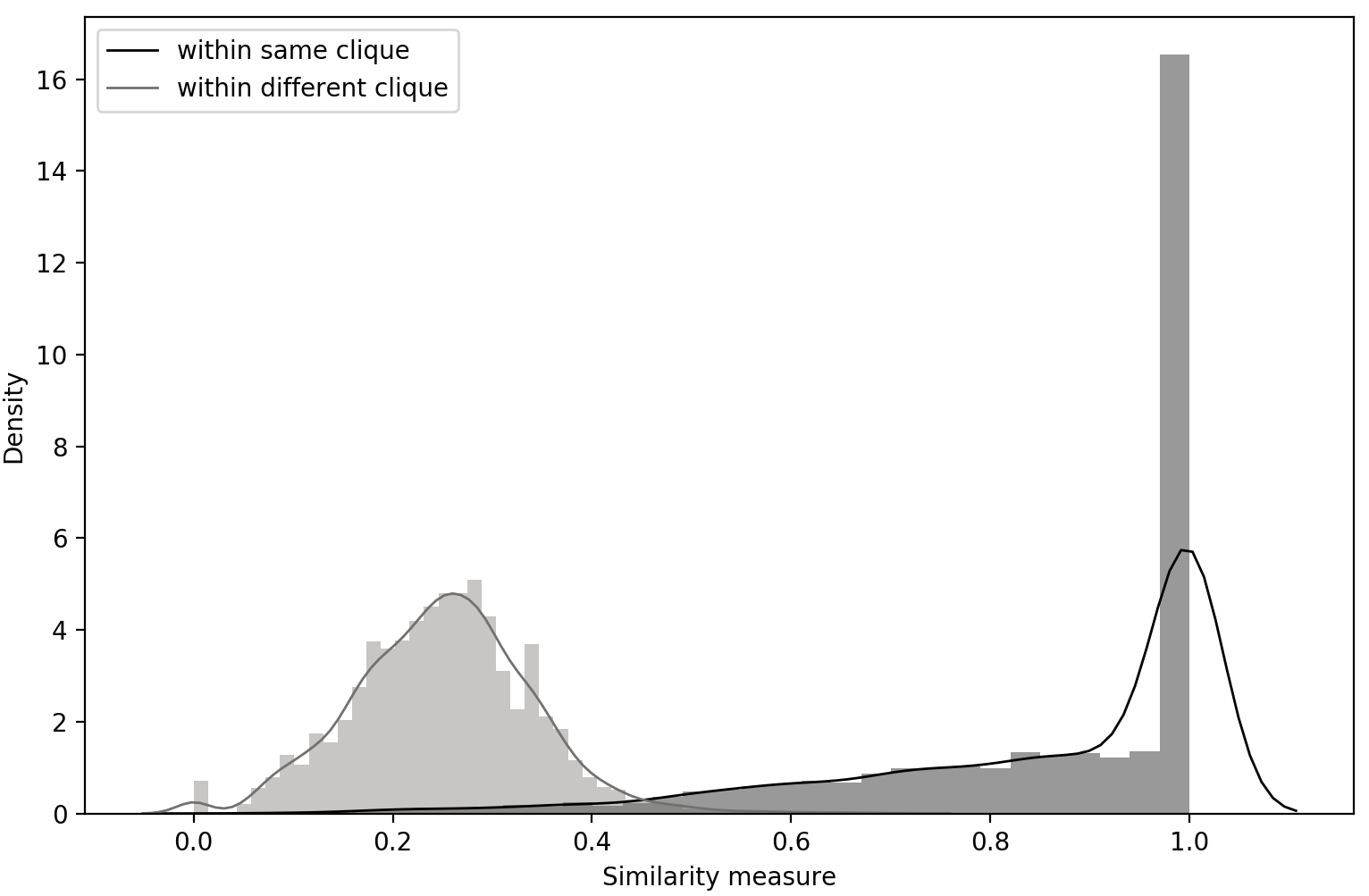}}}
 \caption{String similarity of MSD song titles in the SHS train set within and outside its clique}
 \label{fig:clique_similarity}
\end{figure}

We did a preliminary analysis on understanding how the MSD song titles are similar within songs in the same clique and other cliques. This concept of understanding cover similarity within a cover clique were previously explored through audio in \cite{2009MIREX}. Figure \ref{fig:clique_similarity} shows the distribution of similarity measures for all MSD song titles within and outside their respective cliques in the SHS train set. This similarity measure is computed by the Levenshtein distance ratio\footnote{https://github.com/ztane/python-Levenshtein} between the two song titles. Here 1 indicates maximum similarity and 0 indicates none. From the Figure \ref{fig:clique_similarity}, we can clearly see a strong separability and discriminatory property of song-title in differentiating covers in the case of SHS train set as expected. This can be true for most of the popular music cover versions without language invariance when metadata are properly attributed. For example, the song ``\textit{All Along The Watch Tower}" by Bob Dylan has cover songs with titles ``\textit{All Along The Watchtower (Jimi Hendrix cover)}", ``\textit{ Along The Watchtower (2001 Digital Remaster)}" etc.  But this similarity in titles are not preserved in lot of non-pop music. For example in classical music, a song in a cover clique is often distinguished using unique identifiers like opus numbers while most of the other strings in the title exhibits strong string similarity with songs in other cover cliques (eg. \textit{Chopin Mazurka in C major Op. 7, No. 5} and \textit{Chopin Mazurka in C major Op. 24, No. 2} are two different works of art). From our observation, different non-pop music like western classical music are not well represented in the MSD. This is one of the limitations of evaluating generic cover detection systems on SHS and MSD.

\subsubsection{Language invariance}

One of the main challenges of using song titles and lyrics for cover retrieval is the language variance. To understand the scope of this problem, we detect the language of songs in MSD using their available lyrics or song title (in case of no lyrics). From this process, almost 70 percent of songs in the MSD were detected as in English, thanks to the open-source library \textit{language-detection}\footnote{https://github.com/shuyo/language-detection}. A \textit{unknown} is returned, when no language is detected. From empirical observation we noticed that among the 16{\%} of \textit{unknown} flagged songs, a considerable part of them are actually in English. Beside that, approximately 3{\%} of songs are in Spanish, 2{\%} in French and majority of the remaining in various European languages. Conceptually, a simple way to achieve language invariance would be to translate text into a common language, like English. While this approach should be explored in future studies, we decided to ignore this problem due to the low impact it will have on our mostly-English dataset. Note that this English predominance may be a another drawback of SHS and MSD for evaluating generic text-based cover detection systems.

\subsubsection{Effects of duplicates and unannotated covers}

The presence of duplicates and unannotated SHS covers within MSD are a known problem within the MIR community\footnote{https://labrosa.ee.columbia.edu/millionsong/secondhand}. MSD creators provide an official list of MSD duplicates\footnote{https://labrosa.ee.columbia.edu/millionsong/blog/11-3-15-921810-song-dataset-duplicates}. Particularly in the case of cover song detection, this has been explained in the work of \cite{Osmalskyj2015:08}. This can create two problems while evaluating cover identification systems with SHS and MSD datasets. The presence of duplicates might give overly optimistic results when using audio-based methods since query and reference tracks are sometimes identical and thus yielding a low distance. To address the duplicate problem, we flagged the official MSD duplicate tracks in our database to exclude them in our experiments. The effect of official MSD duplicates in the final accuracy of experiments are further discussed in Section \ref{sec:final-eval}. On the other hand, the presence of unannotated covers can affect the measured performance of a cover detection system even though it detects the actual covers of the query songs. We found several examples of these cases while exploring the SHS train set where there are many unannotated covers with same or very similar metadata and lyrics as of a query song. This means that the evaluations of cover detection systems on SHS and MSD might not exactly reflect the actual accuracy of the proposed system.

\section{Experimental setup}\label{sec:Experiments}

\subsection{Methodology}

In this Section, we detail our proposed experimental system. We introduce a cover detection system which leverages song metadata, lyrics  and audio comprehensively. Regarding evaluation of our system, we followed the same standard approach as in \cite{BertinMahieux2012:01, Humphrey2013:02}, thus making it comparable to existing state of the art large-scale cover detection systems. We report MAP \cite{IRintroductionStandford} at top-\textit{k} results as evaluation metric for our experiments since it reflects the accuracy of covers identified in high ranks. Other metrics such as \textit{mean reciprocal rank}, \textit{mean rank of first cover} and \textit{average rank} are not found to be reliable since we use top-\textit{k} results of the search response. We did our exploration experiments on the SHS train set (12,960 tracks) and kept the SHS test set (5,238 tracks) aside for the final evaluation as in \cite{Humphrey2013:02}.

\begin{table}
 \begin{center}
 \begin{tabular}{|l|l|}
  \hline
   ID & Description \\
  \hline
  title & search by MSD-song-title\\
  \hline
  pre-title & search by pre-processed MSD-song-title\\
  \hline
  mxm-lyr & search by MXM-lyrics\\
  \hline
  + & denotes merging and re-ranking process
  		\\
  \hline
 \end{tabular}
\end{center}
 \caption{Summary of different methods used in our experiments}
 \label{tab:system-summary}
\end{table}

\subsection{Text-based approaches}

In this section, we further explain the methods used to retrieve text-based cover song similarity from song titles and lyrics. Here, similarity of text is obtained by computing the cosine similarity of weighted vectors, where the weighted vectors are the \textit{term frequency(tf)} and \textit{inverted document frequency(idf)} computed from the document. \textit{term frequency} is a count of how many times a word appear in a given bag-of-words document and \textit{inverse document frequency} is the number of times a word occurs in a corpus of documents. We used the standard search and text engine Apache Lucene\footnote{https://lucene.apache.org/} for computing this text similarity. Lucene uses a combination of Boolean Space (BM) and Vector Space Model (VSM) of information retrieval \cite{salton1975VSM} for computing this cosine similarity. Readers are referred to lucene scoring formula\footnote{https://lucene.apache.org/core/4{\_9\_0}/core/org/apache/lucene/search-/similarities/TFIDFSimilarity.html} for further details on the performance optimization functions adapted in Lucene. \textit{tf-idf}-based search is a standard process in information retrieval and well adapted in all modern search engines. Moreover, \textit{tf-idf} methods can be easily adapted to the \textit{bag-of-words} lyrics provided in the MXM dataset. 

\subsubsection{Metadata-based approach}

From the analysis in Section \ref{sec:cover-set-similarity}, we already see that the song titles are a potential discriminatory feature to detect cover songs in the case of SHS train set. Cover songs titles usually have certain keywords like \textit{live, cover, ep, studio, acoustic, remix, remastered, version} etc  to often describe various types of cover. From our exploration on SHS train set we noticed that the presence of these keywords often introduce a variance in the song titles among a cover clique. Hence, we stored a preprocessed MSD title as a new metadata field, removing this invariance. We define a codebook of these common keywords (total 34  keywords) by exploring the SHS train set (eg. \textit{version, remaster, live, edit, album, demo, reprise, edit, radio, original, cover, single etc}). We parsed strings appearing inside parenthesis in MSD song titles. The parsed string is then removed from the song title if a possible match has been found with its stemmed version and any keywords in the defined codebook. This preprocessing step of keywords is very similar to the flagging approach done in \cite{SmithYoutube2017:00}. 11.38 {\%} of songs in the whole MSD and 12.42 {\%} of songs in the SHS train set were affected by this preprocessing step. Our experiments on the SHS train sets achieved better accuracy on pre-processed MSD titles than the original ones, as one can read from Table \ref{tab:shs-train-msd-results}. This shows that removing non-relevant information from song titles might help to increase the accuracy of retrieving covers. For example, searching by the song title "\textit{Be  (1994 Digital Remastered Version)}" using \textit{tf-idf}-based methods leads to noisy results found from the words \textit{1994, digital, remastered} and \textit{version}.

Metadata in real world music databases can be untidy and prone to errors. In these cases, customized text processing strategies can be employed. The artist credits associated with a work of art (e.g. Composer, Lyricist...etc) can also be a relevant metadata for cover detection in some cases. However, this valuable metadata is less likely to be found in a large corpus. For instance, MSD does not contain such metadata. This is something to be investigated in future works. 

\subsubsection{Lyrics-based approach}

In the case of cover songs excluding instrumental versions, lyrics encode certain semantic features about a common work of art. Especially in the case of pop-song covers, the lyric content tends to be the same or very similar between a song and its cover. Lyrics retrieval have been previously studied by means of various NLP based alignment techniques\cite{knees2005Lyrics}. For our experiments, we decided to use a bag-of-words (BOW) \textit{tf-idf}-based retrieval as we explained in the beginning of this section. By this experiment we aim to check whether lyrics helps in some way to retrieve it's cover songs. We used the \textit{more like this}\footnote{https://www.elastic.co/guide/en/elasticsearch/reference/current/query-dsl-mlt-query.html} query method in Elasticsearch (ES)\footnote{https://www.elastic.co/products/elasticsearch} to retrieve songs with similar lyrics to avoid reinventing the wheel with custom methods since we want to have a baseline result. This query method selects the top-\textit{k} terms with highest \textit{tf-idf} to form a disjunctive query of these terms to retrieve similar documents. We finally used the parameters \textit{max{\_}query{\_}terms=12} and \textit{min{\_}query{\_}terms=1} of this method by experimenting with various examples in the SHS train set. On the other hand, the lyrics provided by the MXM dataset are also distributed as BOW format which makes it ideal for us to use MXM lyrics in the experiments. 

\begin{table}
 \begin{center}
 \begin{tabular}{|l|l|l|l|}
  \hline
   MAP@k & k=10 & k=100 \\
  \hline
  title & 0.346  & 0.397 \\
  \hline
  pre-title & 0.35 & 0.401 \\
  \hline
  mxm-lyr & 0.137  & 0.140 \\
  \hline
  title + mxm-lyr & \textbf{0.376} & \textbf{0.433 } \\
  \hline
  pre-title + mxm-lyr & 0.372 & 0.431 \\
  \hline
 \end{tabular}
\end{center}
 \caption{Results of experiments with SHS train set against whole MSD.}
 \label{tab:shs-train-msd-results}
\end{table}

\subsection{Data framework}

\begin{figure}
 \centerline{
 {
 \includegraphics[width=7cm, height=3.3cm]{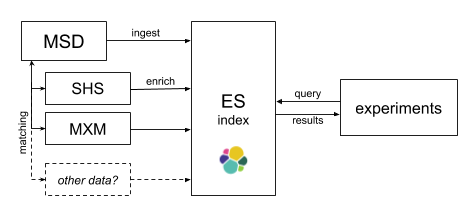}}}
 \caption{Data framework used throughout our experiments with textual data}
 \label{fig:data_framework}
\end{figure}

We use ES as database to store and serve all textual metadata and lyrics used in our large-scale experiments. ES is built on top of the Lucene search engine which makes it an ideal candidate for standard information retrieval and text processing functions. Each track from the MSD was ingested in the ES index, selecting only the metadata used in our experiments. Data from other datasets such as SHS and MXM were then added to existing document, in an update operation. The matching between these datasets and MSD are publicly provided by their respective creators. The overall data framework can be visualized in Figure \ref{fig:data_framework}. During ingestion of textual data, the ES standard analyzer divides text into terms on word boundaries by Unicode Text Segmentation\footnote{http://unicode.org/reports/tr29/} algorithm (this is similar to a stemming process). Readers are referred to ES query-dsl documentation\footnote{https://www.elastic.co/guide/en/elasticsearch/reference/current/query-dsl.html} on further understanding the various search functions available within ES.

\subsection{Audio-based approaches}\label{sec:audio-based}

\begin{figure}
 \centerline{
 {
 \includegraphics[width=8cm,height=5.4cm]{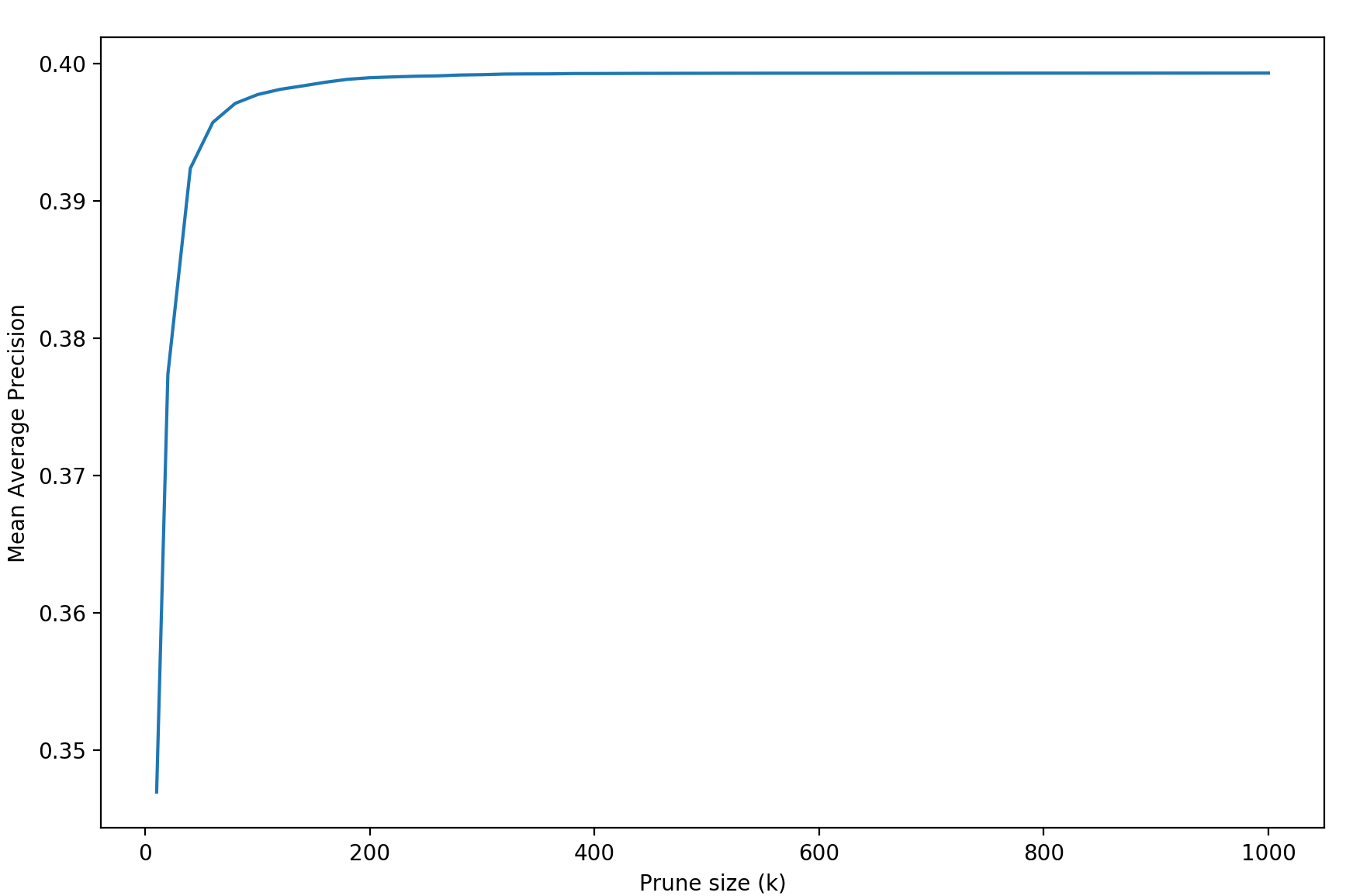}}}
 \caption{MAP@k plot for MSD song title experiment with SHS train set against the whole MSD.}
 \label{fig:map@k_plot}
\end{figure}

The preliminary results in Table \ref{tab:shs-train-msd-results} and Table \ref{tab:shs-train-shs-results} on SHS train set and MSD reveals that metadata-based pruning achieves higher results than the scalable audio-based approaches in the previous works. We decided to use the traditional audio-based cover detection system proposed in \cite{2009MIREX} with a aim to further improve the accuracy achieved by text-based methods: this algorithm achieved the best MAP score (0.75) for this MIREX campaign till date and the executable files are publicly available\footnote{http://joanserra.weebly.com/publications.html}. The output of this algorithm is a distance matrix from pairwise similarities of all the query-reference song pairs in the dataset. A small distance means the reference track is more likely to be a cover of the queried song. A multi-modal framework as proposed in our work allows the evaluation of such a strong audio-based algorithm in large-scale. This means that instead of computing pairwise music similarities of 1 against 999,999 tracks of MSD, we can process pairwise distances of a query against pruned top-\textit{k} results from the text-based methods. We did an analysis to estimate this optimum pruning size for text-based approaches from where we can further apply more specific audio-based methods. As we can see in Figure \ref{fig:map@k_plot}, the MAP metric does not evolve significantly after the top-100 pruning. From this observation, we can safely rely on the top-100 pruned results to further perform audio-based cover detection \cite{2009MIREX} and thus having a trade-off between accuracy and scalability. Effects of this proposed methods are further detailed in Section \ref{sec:final-eval}. 

\subsection{Combining meta-data, lyrics and audio results}\label{sec:mergemethod}

In this section we explain our methods to comprehensively combine the results from song title, lyrics and audio similarity measures. This can be seen as similar to the multi-layer (stage) cover detection frameworks as discussed in \cite{SmithYoutube2017:00} and \cite{Cai2017TwolayerCover}. MAP score increases when actual covers are detected in the top ranks. Thus using content-based features such as lyrics to re-rank the response from song title experiments should increase the MAP of the system. Table \ref{tab:system-summary} shows the summary of different methods we used in the experiments. Here, we obtained the \textit{title + mxm-lyr} merged list by concatenating common documents present in the \textit{title} and \textit{mxm-lyr} results on the top ranks and the rest of documents in the \textit{title} response in the low ranks. Thus we obtain a new merged list \textit{title + mxm-lyr}. By doing so, we minimize some possible errors of the merging process without penalizing the MAP achieved with the \textit{title} method. Note that for some optimization, we only consider the documents with high scores in the \textit{mxm-lyr} results for the above mentioned merging process. These documents are selected by applying a relative threshold \textit{h} on the difference of its score at rank 1 against the score at all other ranks in the \textit{mxm-lyr} results. A default threshold parameter \textit{h}=0.5 was used in our experiments after exploring with various values on the lyrics responses. Table \ref{tab:shs-train-msd-results} shows the results of text-based experiments on SHS train set against the whole MSD. We can see that re-ranking \textit{title} with \textit{mxm-lyr} results boosts the accuracy of \textit{title} method by 3.6{\%} in the case of top 100 results of SHS train set against MSD experiment. Here, we have to take in account that only 23.76{\%} of corresponding lyrics to MSD were available in the MXM dataset. So in this merging process, the system still maintains the accuracy obtained from the \textit{title} method even though there is no lyrical data available for the queried song. This is quite similar to the real world scenarios since music lyrics can be comparatively hard to find in non-commercial music databases. 
Regarding audio methods, our chosen system \cite{2009MIREX} returns a very low distance value (which is close to zero) when the query and reference pairs have high tonal music similarity. Here we obtain the new merged list by re-ranking the best performing text-based method (\textit{title + mxm-lyr}) results by only using the songs with low music similarity distances (as given by \cite{2009MIREX}) to the top ranks. We used \textit{absolute thresholding} to select these songs with low distance scores. A default absolute threshold \textit{h}=$0.1$ was used in our final experiments which was found empirically by exploring the training data. We can see whole this process as a three stage architecture which can be optimally designed using more novel methods in the future works.

\begin{table}
 \begin{center}
 \begin{tabular}{|l|l|}
  \hline
   & MAP  \\
  \hline
  Osmalskyj et. al\cite{Osmalskyj2015:08}  & 0.048 \\
  \hline
  Bertin-Mahieux et. al \cite{BertinMahieux2012:01} & 0.09 \\
  \hline
  Khadkevich et. al\cite{KhadkevichChord2013:07} & 0.10 \\
  \hline
  Humphrey et. al\cite{Humphrey2013:02} & 0.28 \\
  \hline
  title & 0.829 \\
  \hline
  pre-title & \textbf{0.927} \\
  \hline
  mxm-lyr & 0.139 \\
  \hline
  title + mxm-lyr & \textbf{0.921} \\
  \hline
  pre-title + mxm-lyr & \textbf{0.926} \\
  \hline
 \end{tabular}
\end{center}
 \caption{Results of experiments with SHS train set against itself.}
 \label{tab:shs-train-shs-results}
\end{table}

\section{Final evaluation}\label{sec:final-eval}

We now display the results corresponding to the final evaluation of our methods on the unseen data (SHS test set) against the whole MSD. We used the same evaluation methodologies as in \cite{BertinMahieux2012:01} and \cite{Humphrey2013:02}. Table \ref{tab:shs-test-msd-results} compares the MAP scores of all our proposed text-based methods against the previous state of the art works. Note that we report MAP@100 scores of our experiments with and without the official MSD duplicates and MAP@100 is always less than MAP. From Table \ref{tab:shs-test-msd-results}, we can see a 32.5{\%} absolute increase in accuracy by only using MSD song title compared to the current audio-based state of the art \cite{Humphrey2013:02}.This is very interesting considering the fact that song title information is widely and easily available in public compared to other data such as lyrics. Among all, \textit{title + mxm-lyr} method achieved the best MAP score of 0.489 (35.5{\%} absolute increase to \cite{Humphrey2013:02}). Another interesting thing to notice is that the MSD song title preprocessing step explained in Section \ref{sec:Experiments} does not necessarily improve the results on the SHS test set. This shows that more sophisticated preprocessing strategies or text topic-modeling approaches can be borrowed in future works.

Having discussed about the impact of various text-based methods on the experiments with MSD, SHS train and test sets, our next aim is to observe how far the state of the art cover detection algorithms can further improve the accuracy of text-based methods. Table \ref{tab:shs-dzr-results} shows the results of various experiments done on the SHS-DZR test set (4,254 tracks) against MSD-DZR (824,585 tracks) set by excluding all the official MSD duplicates. Note that the baseline results in Table \ref{tab:shs-dzr-results} are computed on the official SHS and MSD datasets, which is slightly different. Confirming this observation and quantifying the results we can see that re-ranking text-based \textit{(title + mxm-lyr)} results using state of the art audio-method \cite{2009MIREX} with low distance thresholds (\textit{h=0.1}) further boosts the MAP score by 2.1{\%}.

\begin{table}
 \begin{center}
 \begin{tabular}{|l|l|l|}
  \hline
    & MAP & MAP (no dupl.) \\
  \hline
  Bertin-Mahieux et al \cite{BertinMahieux2012:01} & 0.029 & -- \\
  \hline
  Humphrey et. al\cite{Humphrey2013:02} & 0.134 & --\\
  \hline
  title  & 0.421 & 0.459 \\
  \hline
  pre-title & 0.421 & 0.453 \\
  \hline
  mxm-lyr & 0.148 & 0.149 \\
  \hline
  title + mxm-lyr & \textbf{0.442} & \textbf{0.489} \\
  \hline
  pre-title + mxm-lyr & \textbf{0.442} & \textbf{0.486} \\
  \hline
 \end{tabular}
\end{center}
 \caption{Results of experiments with SHS test set against whole MSD.}
 \label{tab:shs-test-msd-results}
\end{table}

\begin{table}
 \begin{center}
 \begin{tabular}{|l|l|}
  \hline
   & MAP \\
  \hline
  Bertin-mahieux et al \cite{BertinMahieux2012:01} & 0.029 \\
  \hline
  Humphrey et. al\cite{Humphrey2013:02} & 0.134 \\
  \hline
  title  & 0.478 \\
  \hline
  title + mxm-lyr & 0.510 \\
  \hline
  pre-title + mxm-lyr & 0.506 \\
  \hline
  title + mxm-lyr + Serra et. al \cite{2009MIREX} (\textit{h}=0.1)& \textbf{0.531} \\
  \hline
 \end{tabular}
\end{center}
 \caption{Results of experiments with audio re-ranking on SHS-DZR test set against MSD-DZR}
 \label{tab:shs-dzr-results}
\end{table}

\section{Conclusions}\label{sec:conclusion}

In this study we successfully demonstrated that combining song metadata, lyrics and audio features can significantly improve accuracy and scalability of cover song identification in a simple and easy-to-set-up architecture. We report the best MAP score on the MSD to date with a 35.5{\%} absolute increase with \textit{title + mxm-lyr} method and 32.5{\%} absolute increase using only MSD song titles compared to the current audio content-based state of the art methods. To our knowledge, our work reports the first benchmark results on metadata/lyrics-based cover song detection using SHS, MSD and MXM datasets. 
It is quite important to notice that the results presented in this work addresses the use-case of cover detection in a digital music library where corresponding metadata, lyrics and audio are mostly available such as in the case of music streaming services like Deezer. Hence, our method is not competing with traditional audio-based cover identification methods in the literature when none of the metadata or textual content like lyrics are not available. However, song titles are a widely available music metadata and its high performance as a unique input for cover song identification is indeed promising for stimulating further scientific research.
 To an extent, similarity of song titles and lyrics within cover sets are more interpretable than from audio considering the musicological aspects involved in audio feature computation. Involving text topic-modelling and machine learning models could be explored as an immediate direction for future research on text-based cover detection. 

The main drawback of our method is the loss of some actual covers in the pruned results when cover song pairs have entirely different metadata or lyrical content. In these scenarios, audio-based methods can give better results. On the other hand, considering the case of music metadata catalogs where the audio content is not available, text-based cover song detection methods can be highly beneficial in database structuring, musicological studies etc. Hence, an ideal cover detection system should be able to leverage the strength of multi-modal methods according to the use-case of its applications and thus helping to identify strong generic audio-based cover detection systems. 

Our work is built on top of previous works using public datasets like MSD, SHS and MXM thus making it comparable and easily reproducible (in terms of text-based methods). Finally, for open research and reproducibility all our code and data used for text-based methods are available on-line in a public repository\footnote{https://github.com/deezer/cover{\_}song{\_}detection}.

\section{Acknowledgements}
      The authors would like to thank Dr. Francesco Piccoli for the interesting discussions and reviews during this work.

  The authors would also like to thank everyone in the Deezer R{\&}D team for their valuable comments and helpful suggestions. This work was funded and supported by Deezer S.A, Paris, France. The research leading to this work benefited from the WASABI project supported by the French National Research Agency (contract {\tt ANR-16-CE23-0017-01}).

\bibliography{ISMIRtemplate}

\begin{thebibliography}{10}

\bibitem{BertinMahieux2012:01}
Thierry Bertin-Mahieux and Daniel P~W Ellis.
\newblock {Large-scale cover song recognition using the 2d fourier transform
  magnitude}.
\newblock {\em International Society for Music Information Retrieval}, 2012.

\bibitem{BertinMahieux2011:03}
Thierry Bertin-Mahieux and Daniel~P.W. Ellis.
\newblock {Large-scale cover song recognition using hashed chroma landmarks}.
\newblock {\em IEEE Workshop on Applications of Signal Processing to Audio and
  Acoustics}, 2011.

\bibitem{MSD2011}
Thierry Bertin-Mahieux, Daniel~P.W. Ellis, Brian Whitman, and Paul Lamere.
\newblock The million song dataset.
\newblock In {\em {Proceedings of the 12th International Conference on Music
  Information Retrieval ({ISMIR} 2011)}}.

\bibitem{Cai2017TwolayerCover}
Kang Cai, Deshun Yang, and Xiaoou Chen.
\newblock {Two-layer large-scale cover song identification system based on
  music structure segmentation}.
\newblock {\em 2016 IEEE 18th International Workshop on Multimedia Signal
  Processing, MMSP 2016}, 2017.

\bibitem{canofingerprintreview}
Pedro Cano, Eloi Batle, Ton Kalker, and Jaap Haitsma.
\newblock A review of algorithms for audio fingerprinting.
\newblock In {\em Multimedia Signal Processing, 2002 IEEE Workshop on}, pages
  169--173. IEEE, 2002.

\bibitem{CNNCover2017audio}
Sungkyun Chang, Juheon Lee, Sang~Keun Choe, and Kyogu Lee.
\newblock Audio cover song identification using convolutional neural network.
\newblock {\em arXiv preprint arXiv:1712.00166}, 2017.

\bibitem{XXchen2018fusing}
Ning Chen, Wei Li, and Haidong Xiao.
\newblock Fusing similarity functions for cover song identification.
\newblock {\em Multimedia Tools and Applications}, 77(2):2629--2652, 2018.

\bibitem{Zellis2006beatsync}
Daniel~PW Ellis and Graham~E Poliner.
\newblock Identifyingcover songs' with chroma features and dynamic programming
  beat tracking.
\newblock In {\em Acoustics, Speech and Signal Processing, 2007. ICASSP 2007.
  IEEE International Conference on}. IEEE, 2007.

\bibitem{foster2015identifying}
Peter Foster, Simon Dixon, and Anssi Klapuri.
\newblock Identifying cover songs using information-theoretic measures of
  similarity.
\newblock {\em IEEE/ACM Transactions on Audio, Speech and Language Processing
  (TASLP)}, 23(6):993--1005, 2015.

\bibitem{HeoMetricLearning17}
Hoon Heo, Hyunwoo~J Kim, Wan~Soo Kim, and Kyogu Lee.
\newblock Cover song identification with metric learning using distance as a
  feature.
\newblock 2017.

\bibitem{Humphrey2013:02}
Eric~J. Humphrey, Oriol Nieto, and Juan~P. Bello.
\newblock {Data Driven and Discriminative Projections for Large-scale Cover
  Song Identification}.
\newblock {\em Proc. of the 14th International Society for Music Information
  Retrieval Conference}, 2013.

\bibitem{KhadkevichChord2013:07}
Maksim Khadkevich and Maurizio Omologo.
\newblock {Large-Scale Cover Song Identification Using Chord Profiles}.
\newblock {\em Proceedings of the 14th International Society for Music
  Information Retrieval Conference (ISMIR-2013)}, 2013.

\bibitem{SchedlReview2016music}
Peter Knees and Markus Schedl.
\newblock {\em Music Similarity and Retrieval: An Introduction to Audio-and
  Web-based Strategies}, volume~36.
\newblock Springer, 2016.

\bibitem{knees2005Lyrics}
Peter Knees, Markus Schedl, and Gerhard Widmer.
\newblock Multiple lyrics alignment: Automatic retrieval of song lyrics.
\newblock In {\em International Society for Music Information Retrieval
  Conference (ISMIR)}, 2005.

\bibitem{IRintroductionStandford}
Christopher~D Manning, Prabhakar Raghavan, Hinrich Sch{\"u}tze, et~al.
\newblock {\em Introduction to information retrieval}, volume~1.
\newblock Cambridge university press Cambridge, 2008.

\bibitem{MuellerChromaToolbox_ISMIR}
Meinard M{\"u}ller, Frank Kurth, and Michael Clausen.
\newblock Audio matching via chroma-based statistical features.
\newblock In {\em Proceedings of the 6th International Conference on Music
  Information Retrieval ({ISMIR})}, 2005.

\bibitem{OramasMulti-Label2017}
Sergio Oramas, Oriol Nieto, Francesco Barbieri, and Xavier Serra.
\newblock {Multi-label Music Genre Classification from Audio, Text, and Images
  Using Deep Features}.
\newblock {\em International Conference on Music Information Retrieval
  (ISMIR)}, 2017.

\bibitem{OramasRecSys2017}
Sergio Oramas, Oriol Nieto, Mohamed Sordo, and Xavier Serra.
\newblock {A Deep Multimodal Approach for Cold-start Music Recommendation}.
\newblock {\em Proceedings of the 2nd Workshop on Deep Learning for Recommender
  Systems - DLRS 2017}.

\bibitem{EfficientpruningOsmalski}
J.~Osmalskyj, S.~Piérard, M.~Van Droogenbroeck, and J.~J. Embrechts.
\newblock Efficient database pruning for large-scale cover song recognition.
\newblock In {\em 2013 IEEE International Conference on Acoustics, Speech and
  Signal Processing}, pages 714--718, 2013.

\bibitem{Osmalskyj2015:08}
Julien Osmalskyj, Peter Foster, Simon Dixon, and Jean-jacques Embrechts.
\newblock {Combining features for cover song identification}.
\newblock {\em 16th International Society for Music Information Retrieval
  Conference (ISMIR)}, 2015.

\bibitem{RafiLiveFingerprint}
Z.~Rafii, B.~Coover, and J.~Han.
\newblock An audio fingerprinting system for live version identification using
  image processing techniques.
\newblock In {\em 2014 IEEE International Conference on Acoustics, Speech and
  Signal Processing (ICASSP)}, pages 644--648, 2014.

\bibitem{Ravuri2009}
Suman~V. Ravuri and Daniel P.~W. Ellis.
\newblock Cover song detection: From high scores to general classification.
\newblock In {\em Proceedings of the {IEEE} International Conference on
  Acoustics, Speech and Signal Processing, {ICASSP} 2010, Texas, {USA}}, 2010.

\bibitem{salamon2013tonal}
Justin Salamon, Joan Serra, and Emilia G{\'o}mez.
\newblock Tonal representations for music retrieval: from version
  identification to query-by-humming.
\newblock {\em International Journal of Multimedia Information Retrieval},
  2(1):45--58, 2013.

\bibitem{salton1975VSM}
Gerard Salton, Anita Wong, and Chung-Shu Yang.
\newblock A vector space model for automatic indexing.
\newblock {\em Communications of the ACM}, 18(11):613--620, 1975.

\bibitem{Serra2008:06}
J.~Serra, E.~Gomez, P.~Herrera, and X.~Serra.
\newblock Chroma binary similarity and local alignment applied to cover song
  identification.
\newblock {\em IEEE Transactions on Audio, Speech, and Language Processing},
  2008.

\bibitem{OTISerra}
Joan Serr{\`a}, Emilia G{\'o}mez, and Perfecto Herrera.
\newblock Transposing chroma representations to a common key.
\newblock {\em IEEE CS Conference on The Use of Symbols to Represent Music and
  Multimedia Objects}, 2008.

\bibitem{SerraReview2010}
Joan Serr{\`{a}}, Emilia G{\'{o}}mez, and Perfecto Herrera.
\newblock {Audio cover song identification and similarity: Background,
  approaches, evaluation, and beyond}.
\newblock {\em Studies in Computational Intelligence}, 274, 2010.

\bibitem{Serra2009:05}
Joan Serra, Xavier Serra, and Ralph~G. Andrzejak.
\newblock {Cross recurrence quantification for cover song identification}.
\newblock {\em New Journal of Physics}, 2009.

\bibitem{2009MIREX}
Joan Serr{\`a}, Massimiliano Zanin, and R.~G. Andrzejak.
\newblock Cover song retrieval by cross recurrence quantification and
  unsupervised set detection.
\newblock In {\em Music Information Retrieval Evaluation eXchange (MIREX)
  extended abstract}, 2009.

\bibitem{silva2016simple}
Diego~F Silva, Chin-Chin~M Yeh, Gustavo Enrique de Almeida Prado~Alves Batista,
  Eamonn Keogh, et~al.
\newblock Simple: Assessing music similarity using subsequences joins.
\newblock In {\em International Society for Music Information Retrieval
  Conference}, 2016.

\bibitem{SmithYoutube2017:00}
J.~B.~L. Smith, M.~Hamasaki, and M.~Goto.
\newblock Classifying derivative works with search, text, audio and video
  features.
\newblock In {\em 2017 IEEE International Conference on Multimedia and Expo
  (ICME)}, July 2017.

\bibitem{tralie2017mfcc}
Christopher~J Tralie.
\newblock Early mfcc and hpcp fusion for robust cover song identification.
\newblock {\em International Conference on Music Information Retrieval
  (ISMIR)}, 2017.

\bibitem{KnownArtistLive}
T.~Tsai, T.~Prätzlich, and M.~Müller.
\newblock Known-artist live song identification using audio hashprints.
\newblock {\em IEEE Transactions on Multimedia}, 2017.

\bibitem{van2014cognition}
JMH van Balen, Dimitrios Bountouridis, Frans Wiering, and Remco~C Veltkamp.
\newblock Cognition-inspired descriptors for scalable cover song retrieval.
\newblock In {\em proceedings of the 15th international conference on Music
  Information Retrieval}, 2014.

\end{thebibliography}

\end{document}